\documentclass[%
twocolumn,
10pt,
superscriptaddress,
amsmath,amssymb,
aps
]{revtex4-2}

\usepackage{graphicx}
\usepackage{dcolumn}
\usepackage{bm}
\usepackage[braket]{qcircuit}
\usepackage{graphicx}  

\usepackage{graphicx}
\usepackage{amsmath}
\usepackage{braket}
\usepackage[capitalise]{cleveref}
\makeatletter
\if@cref@capitalise
\Crefname{figure}{Fig.}{Figs.}
\Crefname{section}{Section}{Sections}
\Crefname{subsection}{Section}{Sections}
\Crefname{subsubsection}{Section}{Sections}
\else
\crefname{figure}{fig.}{figs.}
\crefname{section}{section}{sections}
\crefname{subsection}{section}{sections}
\crefname{subsubsection}{section}{sections}
\fi
\makeatother

\usepackage{subfig} 

\usepackage{lipsum}
\usepackage{xcolor} 
\usepackage{nicefrac}

\newcommand{\qcrank}{\textbf{\texttt{QCrank}}}

\newcommand{\eh}{\textbf{\texttt{EHands}}}

\newcommand{\ehp}{\textbf{\texttt{EHands protocol}}}
\newcommand{\moabs}{\textbf{\textbf{Monarq}}}
\newcommand{\mo}{\textbf{\texttt{Monarq}}}

\usepackage{tikz}  

\begin{document}
\preprint{APS/123-QED}

\title{Sequence and Image Transformations with \moabs: \\Quantum Implementations for NISQ  Devices}

\author{Jan Balewski}
\affiliation{National Energy Research Scientific Computing Center, Lawrence Berkeley National Laboratory, Berkeley, CA, USA}
\author{Roel Van Beeumen}
\email{Corresponding authors: tperciano@lbl.gov, rvanbeeumen@lbl.gov}
\affiliation{Applied Mathematics and Computational Research Division, Lawrence Berkeley National Laboratory, Berkeley, CA, USA}
\author{E. Wes Bethel}
\affiliation{Computer Science Department, San Francisco State University, San Francisco, CA, USA}
\author{Talita Perciano}
\email{Corresponding authors: tperciano@lbl.gov, rvanbeeumen@lbl.gov}
\affiliation{Scientific Data Division, Lawrence Berkeley National Laboratory, Berkeley, CA, USA}

\begin{abstract}
We introduce \mo~ -- a unified quantum data processing framework combining \qcrank\ encoding with \ehp\ for polynomial transformations -- and
demonstrate its implementation on noisy intermediate-scale quantum (NISQ)
hardware.
This framework provides fundamental quantum building blocks for signal and image processing tasks---including convolution, discrete-time Fourier transform (DFT), squared gradient computation, and edge detection---serving as a reference  for a broad class of data processing applications on near-term quantum devices.

\end{abstract}

\maketitle

\section{Introduction}
\label{sec:intro}

Quantum computing presents new possibilities for signal and image processing tasks~\cite{Yan2020QIP,Yao2017QIP,10.1145/3663577}. However, translating theoretical quantum algorithms into practical applications on {\it noisy intermediate scale quantum} (NISQ) devices poses significant challenges due to inherent noise, limited coherence times, and restricted qubit connectivity~\cite{RevModPhys.94.015004,Cerezo2021VQA,gonzalez2022error}.

This work addresses the question: \textbf{Can quantum implementations of classical data processing operations  yield reliable results on current NISQ hardware?} We focus on shallow circuits tailored for NISQ constraints with the aim of introducing new building blocks for quantum  operations useful for developing future complex quantum algorithms but without establishing quantum advantage for the building blocks over classical methods.

Key operations in signal and image processing—such as convolution, Fourier transforms, and edge detection—are critical across applications from medical imaging to autonomous systems~\cite{Wang2022QIPreview,10.1145/3663577,gonzalez2018digital}. Although these operations are effectively performed on classical hardware, understanding their quantum counterparts can illuminate the potential of quantum computing in data processing. Our objective is to develop effective quantum implementations of these operations as foundational elements for advanced algorithms.

Despite existing theoretical proposals for these operations, practical implementations have faced hurdles, particularly: (1) efficiently encoding data into quantum states and (2) performing calculations before decoherence. This highlights the need for quantum implementations tailored for NISQ constraints~\cite{ebadi2022quantum,lienhard2023rydberg}.

Recent advances in encoding and processing have begun to overcome these challenges. The \qcrank\ encoding scheme~\cite{qbart} enables efficient input of real-valued sequences into quantum processors using parallel gate structures. Meanwhile, the \ehp~\cite{balewski2025ehands} framework facilitates polynomial transformations on quantum hardware through elementary operations, making it suitable for NISQ environments.

The contributions of this work are threefold:
First, we introduce the framework for integrating \eh\ polynomial computation with \qcrank\ data encoding via a shared EVEN encoding scheme to create a streamlined quantum data processing pipeline.
Second, we validate four diverse algorithms, establishing \textbf{benchmarks} for the framework: convolution and squared gradient computation on IBM quantum hardware, and discrete-time Fourier transform and edge detection using ideal simulators for larger problem sizes.
Third, we propose a methodology for assessing processing accuracy by comparing root mean square error (RMSE) against known outcomes, showing that meaningful computational results are achievable despite current hardware limitations.
We refer to this integrated pipeline as the \mo~ framework, detailed in
Section~\ref{sec:framework}.

The remainder of this paper is organized as follows. Section~\ref{sec:related} reviews related work on quantum data representations and polynomial transformations, and provides technical background on the \qcrank\ and \eh\ protocols. Section~\ref{sec:framework} details the \qcrank+\eh\ framework, its integration through EVEN encoding, and circuit constructions for four applications. Section~\ref{sec:results} presents our experimental methodology and results on IBM quantum hardware and ideal simulators. Finally, Section~\ref{sec:conclusion} discusses current limitations and future research directions.
%
%
%

\section{Related Work and Background}
\label{sec:related}

This section reviews prior work in quantum data representation and processing, and establishes the technical foundations for \qcrank\ encoding and \eh\ polynomial transformations that underpin our framework.


\subsection{Quantum Data Representations and Encoding}
\label{sec:related_encoding}
Quantum image processing (QIMP) was introduced in 2003 by Venegas-Andraca and Bose~\cite{VenegasAndraca2003QIMP} with the {\it qubit lattice model}. Several quantum image representations followed: the {\it flexible representation of quantum images} (FRQI)~\cite{LePQ2011FRQI} encodes pixel values as rotation angles on a single qubit, while the {\it novel enhanced quantum representation} (NEQR)~\cite{Zhang2013NEQR} encodes pixel values in the computational basis, enabling arithmetic operations on binary values. Extensions include {\it multi-channel representation} (MCRQI)~\cite{Sun2013MCRQI} for color images with reduced circuit complexity~\cite{Khan2019IFRQI}. The QPIXL framework~\cite{Amankwah2022QuantumImages} unified these representations using uniformly controlled rotation (UCR) gates.

\paragraph{\qcrank\ encoding.}
The \qcrank\ encoding scheme~\cite{qbart} builds upon these foundations by introducing parallel UCR gates that significantly reduce circuit depth. \qcrank\ uses two sets of qubits: $n_a$ \emph{address qubits} and $n_d$ \emph{data qubits} (channels). We use index $i \in \{0, 1, \ldots, 2^{n_a}-1\}$ to denote the address , and index $j \in \{0, 1, \ldots, n_d-1\}$ to denote the data channel. The real value at address $i$, channel $j$ is denoted $x_{i,j} \in [-1, 1]$. In total, \qcrank\ encodes $n_d \cdot 2^{n_a}$ real values using $n_a + n_d$ qubits.

For one-dimensional sequences of length \( L \), we typically employ \( n_d = 1 \) data qubit and \( n_a = \lceil \log_2 L \rceil \) address qubits, where each sequence element \( x_i \) is stored at address \( i \). In scenarios that require processing multiple related sequences—such as a signal and its modulations—multiple data channels with \( n_d > 1 \) can be utilized to store different sequences at the same addresses.


\qcrank\ prepares the quantum state:
\begin{equation}
\ket{\psi} = \frac{1}{\sqrt{2^{n_a}}} \sum_{i=0}^{2^{n_a}-1} \ket{i} \otimes \left( \bigotimes_{j=0}^{n_d-1} \ket{\sqrt{1+x_{i,j}}} \right),
\label{eq:qcrank_state}
\end{equation}
where $\ket{i}$ denotes the computational basis state of the address qubits corresponding to index $i$ (the binary representation of integer $i$), and $\ket{\sqrt{1+x_{i,j}}}$ is the encoded state of the $j$-th data qubit. Each $x_{i,j}$ value is encoded via rotation $R_y(\theta_{i,j})$ where $\theta_{i,j} = \arccos(x_{i,j})$, following the expectation value encoding (EVEN) scheme~\cite{balewski2025ehands}. Data recovery relies on measurement of expectation value of Pauli-Z operator:
\begin{equation}
x_{i,j}^{\text{meas}} = \langle \sigma_z^{(j)} \rangle_i = 1 - 2p_{j|i},
\label{eq:data_recovery}
\end{equation}
where $p_{j|i}$ is the probability of measuring the $j$-th data qubit in state $\ket{1}$ conditioned on address $\ket{i}$.

For 2D images of size $W \times H$, pixel coordinates $(a, b)$ map to linear index $i = a \cdot W + b$. The parallel UCR implementation reduces CNOT depth by up to a factor of $n_d$ compared to serial implementations, making \qcrank\ well-suited for NISQ devices. Further details, including the Walsh-Hadamard transformation for angle decomposition, can be found in~\cite{qbart}.


\subsection{Quantum Processing and Polynomial Transformations}
\label{sec:related_processing}

Processing data encoded in quantum states requires operations that respect the
constraints of near-term hardware.  
We review three relevant lines of work before introducing the \ehp\ that underpins our approach.

\paragraph{Quantum image processing and convolution.}
Few quantum image processing algorithms have been experimentally demonstrated.
Yao et al.~\cite{Yao2017QIP} implemented quantum edge detection on NMR hardware
using Hadamard transforms, while proposals for quantum
filtering~\cite{Jiang2019QFilter}, image
transformation~\cite{Caraiman2015QTransform}, and Prewitt operator
implementations~\cite{Zhou2019QuantumEdge} remain theoretical due to deep
circuits and connectivity constraints incompatible with NISQ devices.  Closely
related is the {\it quantum Fourier transform} (QFT), a key subroutine in
algorithms such as Shor's factoring~\cite{Shor1994}, with small-scale
experimental demonstrations~\cite{Weinstein2001QFT, Corcoles2021IBMQFT}, though
scaling remains challenging.  Quantum convolution has been explored in quantum
neural networks~\cite{Cong2019, Henderson2020} and signal
processing~\cite{Grigoryan2022QuantumConvolution}, typically leveraging
interference effects, but practical hardware implementations have been limited.

\paragraph{Polynomial transformation frameworks.}
A common thread in these applications is the need for polynomial transformations
of encoded data.  {\it Quantum signal processing} (QSP)~\cite{Low2017QSP}
implements such transformations through rotation sequences, and {\it quantum
singular value transformation} (QSVT)~\cite{Gilyen2019QSVT} extends this to
block-encoded matrices.  While theoretically powerful, both techniques require
deep circuits that are challenging for current hardware.  Variational
approaches~\cite{Cerezo2021VQA, Benedetti2019VQE} offer shallower alternatives
but introduce barren plateau issues~\cite{McClean2018BarrenPlateaus}.

\paragraph{\eh\ protocol.}
The \ehp~\cite{balewski2025ehands} takes a different approach, defining elementary arithmetic operations that compose into polynomial transformations with shallow circuits. \eh\ uses the same EVEN encoding as \qcrank, where a real value $x \in [-1,1]$ is encoded as:
\begin{equation}
\ket{\sqrt{1+x}} = R_y(\theta)\ket{0}, \quad \text{where } \theta = \arccos(x).
\label{eq:even}
\end{equation}

\begin{figure}[h!]
\hspace{0.1mm}
\Qcircuit @C=1em @R=1em {
&~~~~~~~ \Pi : ~~\text{Product-with-memory (multiplication)} \\
  \rightarrow &&   \qw & \ctrl{1} & \qw & \\
\rightarrow & & \gate{R_z(\frac{\pi}{2})} & \targ & \qw & \rightarrow\\
}\hspace{5mm}
~\\ \vspace{3mm}
\hspace{1mm}
\Qcircuit @C=1em @R=1em {
&&\Sigma :~~\text{Weighted sum (addition)} \\
 \rightarrow &  & \qw & \ctrl{1} & \gate{R_y(\frac{\alpha}{2})} & \targ & \gate{R_y(-\frac{\alpha}{2})} & \qw  &\rightarrow\\
 \rightarrow &&  \gate{R_z(\frac{\pi}{2})} & \targ & \qw & \ctrl{-1} & \qw 
}
\caption{\textbf{EHands primitives.} Quantum circuits  for   multiplication ($\Pi$) and  addition ($\Sigma$) of 2 real numbers.}
\label{fig:EHands-ops}
\end{figure}


\begin{figure*}[bht!]
\centering
\[
\hspace{12mm}\Qcircuit @C=.4em @R=.3em {
&&& &&&  &&&& \text{QCrank encoding}  &&  &&&&&&& &&& &&  \Delta_i=(I_{i+1}-I_{i-1})/2    &&&&  ~~~\Delta_{i}^2 &\\
\lstick{addr_0 } &  \gate{H}  & \qw & \ctrl{3} & \qw & \ctrl{5} & \qw & \ctrl{2} & \qw & \ctrl{4} & \qw & \qw & \qw & \ctrl{3} & \qw & \ctrl{5} & \qw & \ctrl{2} & \qw & \ctrl{4} & \qw& \barrier[-2.1em]{5} \qw &  \qw & \qw & \qw &  \qw & \qw &  \barrier[-2.em]{5}\qw &  \qw & \qw&  \barrier[-1.4em]{5}  \qw  &\meter \\
\lstick{addr_1 } & \gate{H}  & \ctrl{1} & \qw & \ctrl{3} & \qw & \qw & \qw & \ctrl{2} & \qw & \ctrl{4} & \qw & \ctrl{1} & \qw & \ctrl{3} & \qw & \qw & \qw & \ctrl{2} & \qw & \ctrl{4} & \qw & \qw &  \qw & \qw & \qw &  \qw & \qw &\qw & \qw &\qw &\meter \\
 \lstick{\{I_{i-1}\}}    & \gate{R_y} & \targ & \qw & \qw & \qw & \gate{R_y} & \targ & \qw & \qw & \qw & \gate{R_y} & \targ & \qw & \qw & \qw & \gate{R_y} & \targ & \qw & \qw & \qw & \qw & \gate{X}      & \ctrl{1} & \gate{R_y(\frac{\pi}{4})} & \targ& \gate{R_y(\frac{-\pi}{4})} & \qw  &\qw   & \ctrl{2} & \qw& \measuretab{Z}\\
 \lstick{\{I_{i+1}\}}   & \gate{R_y} & \qw & \targ & \qw & \qw & \gate{R_y} & \qw & \targ & \qw & \qw & \gate{R_y} & \qw & \targ & \qw & \qw & \gate{R_y} & \qw & \targ & \qw & \qw & \qw& \gate{R_z(\frac{\pi}{2})} &  \targ &\qw  & \ctrl{-1} &  \qw &  & &&&\\
  \lstick{\{I_{i-1}\}}   & \gate{R_y} & \qw & \qw & \targ & \qw & \gate{R_y} & \qw & \qw & \targ & \qw & \gate{R_y} & \qw & \qw & \targ & \qw & \gate{R_y} & \qw & \qw & \targ & \qw & \qw&  \gate{X}      & \ctrl{1} & \gate{R_y(\frac{\pi}{4})} & \targ& \gate{R_y(\frac{-\pi}{4})} & \qw&\gate{R_z(\frac{\pi}{2})} &  \targ& \qw &\qw\\
 \lstick{\{I_{i+1}\}}    & \gate{R_y} & \qw & \qw & \qw & \targ & \gate{R_y} & \qw & \qw & \qw & \targ & \gate{R_y} & \qw & \qw & \qw & \targ & \gate{R_y} & \qw & \qw & \qw & \targ& \qw & \gate{R_z(\frac{\pi}{2})} &  \targ &\qw  & \ctrl{-1} &  \qw &  &  &&&\\\
}
\]
\caption{\textbf{Example circuit computing the squared gradient on a sequence of 4 real values $I_i \in [-1,1]$.} \qcrank\ encodes 4 copies of the input sequence on 6 qubits. \eh\ negation and weighted sum are computed in 2 copies, which are then multiplied using the \eh\ product operator. The resulting $\Delta_i^2=\nicefrac{(I_{i+1}-I_{i-1})^2 }{4}$ is measured as an expectation value on the 3rd qubit, conditioned on the bitstring measured on the first two qubits, providing the address $i \in [0,3]$.}
\label{fig:circ_sqrGrad}
\end{figure*}
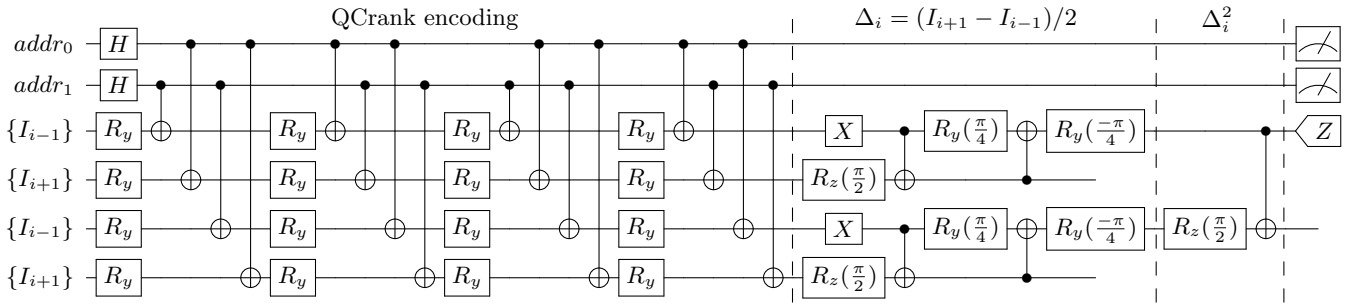

The \eh~ fundamental operations are shown in Fig.~\ref{fig:EHands-ops}. The  {\bf product-with-memory} operation ($\Pi$) multiplies two real values using a single CNOT gate and result is carried by the bottom qubit. The {\bf weighted sum }operation ($\Sigma$) computes $wx + (1-w)y$ using 2 CNOTs and result is on the top qubit. {\bf Negation} (not shown) transforms $x \to -x$ via an X-gate. These operations compose to compute degree-$d$ polynomials with linear scaling: $3d$ qubits, $5d-2$ CNOT gates, and circuit depth $4d$---well-suited for NISQ constraints. Complete details are provided in~\cite{balewski2025ehands}.
 
\Cref{fig:circ_sqrGrad} illustrates how \qcrank\ and \eh\ combine into a complete data processing circuit. This example computes the squared gradient on a sequence---a quadratic polynomial in two variables. \qcrank\ encodes 4 copies of the 4 data values on 2 address and 4 data qubits, then \eh\ operators (negation, weighted sum, and product) transform the encoded values into the desired polynomial. The result is measured as an expectation value on one data qubit, conditioned on the address measured as a bitstring on 2 address qubits. This concrete example demonstrates the end-to-end pipeline that we generalize in Section~\ref{sec:framework}.


\subsection{Gaps Addressed by This Work}
\label{sec:gaps}

While numerous theoretical proposals exist for quantum image and signal processing, experimental demonstrations on real quantum hardware remain limited. Several gaps persist in the field. First, many quantum algorithms for data processing require deep circuits with numerous controlled operations, making them incompatible with NISQ constraints. Second, most experimental work has focused on isolated components rather than complete end-to-end pipelines. Third, there has been limited integration of efficient encoding schemes with practical processing operations, particularly for polynomial transformations on real-valued data. Fourth, a gap persists between efficient data encoding methods and quantum operators that produce desired computational outcomes on real hardware.

Our work addresses these limitations by combining the \ehp\ for polynomial transformations with the \qcrank\ encoding scheme. The shared EVEN encoding enables seamless integration: data encoded by \qcrank\ is processed by \eh\ operators without conversion. Using this NISQ-optimized approach, we implement convolution and squared gradient computation on IBM quantum hardware, and validate DFT and edge detection on quantum simulators, demonstrating both current capabilities and near-term scalability.

%
%
%
%

\section{\moabs\ Framework}
\label{sec:framework}

This section presents the main contribution of this work: a unified framework, \mo, that integrates \qcrank\ encoding with \eh\ polynomial transformations for quantum data processing. We first describe how these two protocols combine into a complete pipeline, then present circuit constructions for four selected data processing applications.


\subsection{Integration of Encoding and Computation}
\label{sec:integration}

The integration of \qcrank\ and \eh\ creates a complete quantum data processing pipeline. The key insight enabling this integration is that both protocols operate on the same EVEN encoding scheme: \qcrank\ prepares quantum states where classical values $x \in [-1,1]$ are encoded via $R_y$ rotations (Eq.~\ref{eq:qcrank_state}), and \eh\ operations preserve this encoding while computing polynomial transformations (Eq.~\ref{eq:even}). This compatibility means that data encoded by \qcrank\ can be directly processed by \eh\ operators without intermediate conversion steps.


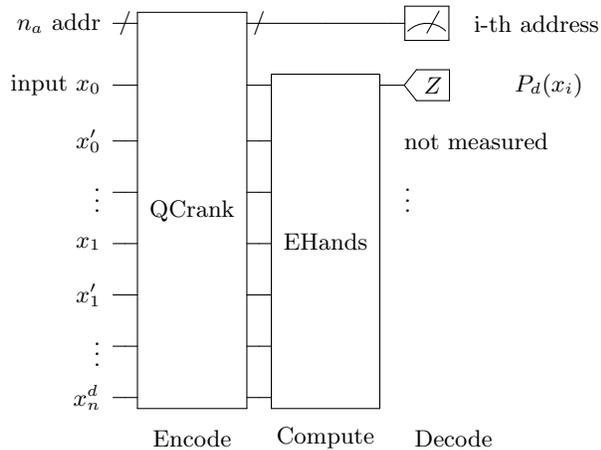
\begin{figure}[t!]
\[
\centering
\Qcircuit @C=.5em @R=1.2em {
\lstick{n_a~ \text{addr}} &  \qw /&\multigate{7}{\text{QCrank}} & \qw /& \qw  & \qw &\meter & \rstick{\text{i-th address}}  \\
\lstick{\text{input } x_0} & \qw & \ghost{\text{QCrank}} & \qw & \multigate{6}{\text{EHands}} & \qw &\measuretab{Z}  & \rstick{~~~~~P_d(x_i)} \\
\lstick{ x'_0}& \qw  & \ghost{\text{QCrank}} & \qw & \ghost{\text{EHands}} &  \rstick{\text{not measured}} \\
\lstick{\vdots} & \qw & \ghost{\text{QCrank}} & \qw & \ghost{\text{EHands}}  & \rstick{\vdots}\\
\lstick{ x_1}& \qw  & \ghost{\text{QCrank}} & \qw & \ghost{\text{EHands}}  \\
\lstick{ x'_1}& \qw  & \ghost{\text{QCrank}} & \qw & \ghost{\text{EHands}}  \\
\lstick{\vdots} & \qw & \ghost{\text{QCrank}} & \qw & \ghost{\text{EHands}}   \\
\lstick{ x_n^{d}}& \qw  & \ghost{\text{QCrank}} & \qw & \ghost{\text{EHands}}  \\
& &\text{Encode} & & \text{Compute} & & & \text{Decode~~} &
}
\]
\caption{\textbf{The \mo{} framework is an integration of \qcrank\ encoding with \eh\ polynomial computation.} For a degree-$d$ polynomial, $d$ copies of each input values $x_i$ are encoded on data qubits. \eh\ operations compute $P_d(x_i)$, retrieved via Pauli-$Z$ measurement one one data qubit conditioned on measured address qubits.}
\label{fig:qcrank-ehands-complete}
\end{figure}

\Cref{fig:qcrank-ehands-complete} illustrates the general structure of the \mo{} framework. The $n_a$ address qubits provide $2^{n_a}$ memory locations indexed by $i$, while the data qubits store encoded values at each address. For computing a degree-$d$ polynomial, the framework requires $d$ copies of each input value $x_i$, encoded across $d$ data qubits at each address. \qcrank\ stores these copies efficiently through its parallel structure, while \eh\ supplies the computational primitives---multiplication, addition, and negation---that compose to evaluate the polynomial. The results are retrieved via measurement of expectation value
of Pauli-Z operator on the output data qubit, conditioned on the address register state $\ket{i}$. A concrete instantiation of this general structure was shown in~\cref{fig:circ_sqrGrad} for the squared gradient computation.

The data flow proceeds in three stages. First, classical input data---either 1D sequences or 2D images with values normalized to $[-1,1]$---is encoded into a quantum state using \qcrank's parallel UCR gates. Next, \eh\ operations act on the data qubits to perform the desired polynomial transformation. Since \eh\ operations are acting only by the data qubit states and not the address qubits, the same transformation $P_d(x_i)$ is applied simultaneously to all $2^{n_a}$ data values, achieving quantum parallelism over the encoded dataset. Finally, measurement of the output data qubit, conditioned on the address register state $\ket{i}$, yields $P_d(x_i)$ for each input value.

A critical feature of this integration is the ability to perform \emph{in-situ summation} by choosing not to measure the address qubits. When address qubits are left unmeasured, the expectation value on the output data qubit automatically yields the average $\frac{1}{2^{n_a}}\sum_i P_d(x_i)$ over all addresses---effectively computing an integral or inner product without explicit accumulation. This capability is exploited in our DFT implementation (Section~\ref{subsubsec:circ_DFT}).


\subsection{Circuit Constructions for Data Processing}
\label{sec:circuits}

This section presents circuit constructions for four applications: convolution, discrete-time Fourier transform, squared gradient computation, and edge detection. Each follows the encode-compute-decode pattern illustrated in~\cref{fig:qcrank-ehands-complete}, but as complexity of the problems grows, so does the size of the circuit implementing it. The results will be discussed in Section~\ref{sec:results}.


\subsubsection{Convolution of Two Sequences}
\label{sec:circ_conv}

Element-wise multiplication of two sequences---the core operation in convolution---corresponds to a degree-1 polynomial in two variables. Given two input sequences $f_i$ and $g_i$ of length $L = 2^{n_a}$, we compute the point-wise product $(f \cdot g)_i = f_i \cdot g_i$ for all $i$ simultaneously.

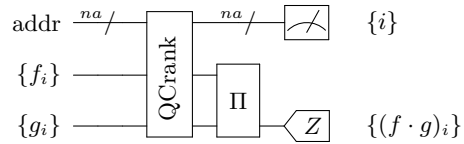
\begin{figure}[h!]
\centering
\hspace{0.1mm}
\Qcircuit @C=1em @R=1em {
   & \lstick{ \text{addr}}  & ^{na}{/} \qw & \qw &  \multigate{2}{\rotatebox{90}{QCrank }}     & ^{na}{/} \qw    & \meter&    \rstick{ \{{i}\}} \\
    &\lstick{\{f_i\}}       &  \qw & \qw & \ghost{\rotatebox{90}{QCrank }}        & \multigate{1}{\Pi}  \\
    &\lstick{\{g_i\}}       &  \qw & \qw & \ghost{\rotatebox{90}{QCrank }}        & \ghost{\Pi}     &  \measuretab{Z}  &  \rstick{ \{(f\cdot g)_i\}} && \\
}
\caption{\textbf{Quantum circuit used for convolution of two lists of length $L = 2^{n_a}$}. The multiplication operator $\Pi$ computes the element-wise product.}
\label{fig:circ-conv}
\end{figure}

The circuit (\cref{fig:circ-conv}) uses \qcrank\ to store both input sequences on $n_a + 2$ qubits: $n_a$ address qubits index the $L$ positions, while 2 data qubits hold $f_i$ and $g_i$ at each address. The \eh\ multiplication operator $\Pi$ computes the product. Measuring the expectation value of the output (bottom) qubit, conditioned on the address register, yields $(f \cdot g)_i$ for each index $i$.

\subsubsection{Discrete-Time Fourier Transform}
\label{subsubsec:circ_DFT}

The DFT demonstrates how integral of convolution can be computed on a QPU by not measuring address qubits, enabling in-situ summation. For a real-valued discrete-time signal $x[n] \in \mathbb{R}$, the DFT is defined by
\begin{equation}
X(\omega) = \sum_{n=-\infty}^{\infty} x[n] e^{-j\omega n}, \quad -\pi \leq \omega < \pi.
\label{eq:DFT}
\end{equation}

\begin{figure*}[ht!]
\centering
\vspace{1cm}
\hspace{0.1mm}
\Qcircuit @C=1em @R=1em {
   & \lstick{\text{addr}}  & ^{na}{/} \qw &   \multigate{6}{\rotatebox{90}{QCrank }}     &  ^{na}{/} \qw    & \qw&\rstick{\text{addresses not measured}} \\
    &\lstick{\{sin(\omega_k t_i)\}}       &  \qw &  \ghost{\rotatebox{90}{QCrank }} &\qw &\qw&\qw &\qw &\qw &\qw  & \multigate{5}{\Pi}   &  \measuretab{Z}    &\rstick{I(\omega_k) }   \\
    &\lstick{\{cos(\omega_k t_i)\}}       &  \qw &  \ghost{\rotatebox{90}{QCrank }}    &\qw &\qw &\qw &\qw    & \multigate{4}{\Pi}    &  \measuretab{Z} & &    & \rstick{Q(\omega_k) }\\
   & \lstick{\vdots}   &  \qw &  \ghost{\rotatebox{90}{QCrank }} & \qw & \vdots &  & &   && &   & \rstick{\vdots}  \\
    &\lstick{\{sin(\omega_1 t_i)\}}       &  \qw &  \ghost{\rotatebox{90}{QCrank }} &\qw  &\qw  & \multigate{2}{\Pi}   &  \measuretab{Z}  &&&&& \rstick{ Q(\omega_1)}\\
    &\lstick{\{cos(\omega_1 t_i)\}}       &  \qw &  \ghost{\rotatebox{90}{QCrank }}       & \multigate{1}{\Pi}   &  \measuretab{Z} &  && & & & & \rstick{I(\omega_1)}  \\
    &\lstick{\{h(t_i)\}}       &  \qw &  \ghost{\rotatebox{90}{QCrank }}        & \ghost{\Pi}  & \qw     & \ghost{\Pi} & \qw  &\ghost{\Pi} & \qw  &\ghost{\Pi} \\
}
\caption{\textbf{Quantum circuit used for DFT}. The input signal $h(t_i)$ of length  $L=2^{n_a}$ as well as $2k$ sequences of trigonometric function are encoded using  \qcrank\ on $n_a+2k+1$ qubits. Next, $2k$   \eh\ multipliers compute $k$ real and $k$ imaginary components. The $n_a$ address qubits are not measured.}
\label{fig:circ-DFT}
\end{figure*}
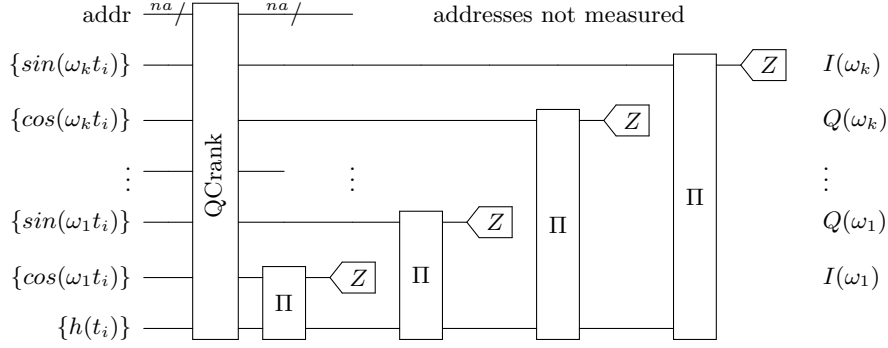

Using the identity $e^{-j\omega n} = \cos(\omega n) - j\sin(\omega n)$, the spectrum decomposes into two real functions:
\begin{equation}
X(\omega) = I(\omega) + jQ(\omega),
\label{eq:DFT_iq}
\end{equation}
where the in-phase and quadrature components are
\begin{align}
I(\omega) &= \sum_n x[n] \cos(\omega n), \label{eq:DFT_i} \\
Q(\omega) &= -\sum_n x[n] \sin(\omega n). \label{eq:DFT_q}
\end{align}
The magnitude and phase follow from $|X(\omega)| = \sqrt{I^2(\omega) + Q^2(\omega)}$ and $\angle X(\omega) = \text{atan2}(Q(\omega), I(\omega))$.
Unlike the standard DFT, which evaluates the spectrum at fixed frequencies $\omega_k = 2\pi k/N$, our approach computes the DTFT at user-selected frequencies, providing flexibility to probe specific spectral regions of interest.

For a finite set of angular frequencies $\{\omega_1, \ldots, \omega_k\}$, the circuit (\cref{fig:circ-DFT}) comprises $n_a + 2k + 1$ qubits. \qcrank\ encodes the input signal $h(t_i)$ of length $L = 2^{n_a}$ along with $2k$ sequences of trigonometric modulations $\{\cos(\omega_m t_i), \sin(\omega_m t_i)\}$. The $2k$ \eh\ multipliers compute element-wise products, and by not measuring the address qubits, the QPU performs the summation in-situ. The circuit yields all $k$ pairs $\{I(\omega_m), Q(\omega_m)\}$, which classical post-processing converts to spectral magnitude and phase.


\subsubsection{Squared Gradient via Central Differences}
\label{sec:circ_grad}

Computing image gradients is fundamental to computer vision and edge detection. The squared horizontal gradient of a 2D grayscale image $I(i,j)$ using central differences is:
\begin{equation}
G_x^2(i,j) = \left( \frac{I(i+1,j) - I(i-1,j)}{2} \right)^2.
\label{eq:sq_grad}
\end{equation}

This operation corresponds to a degree-2 polynomial in two variables, constructed using \eh\ negation, addition, and multiplication operators.

\begin{figure}[h!]
\centering
\hspace{0.1mm}
\Qcircuit @C=1em @R=1em {
    \lstick{\text{addr}}  & ^{na}{/} \qw &   \multigate{4}{\rotatebox{90}{QCrank }}  &\qw   &\qw    & ^{na}{/} \qw    & \meter &  \rstick{ \{{i}\}}\\
    \lstick{\{I_{i-1}\}}       &  \qw & \ghost{\rotatebox{90}{QCrank }}        &  \gate{X}  & \multigate{1}{\sum_{\frac{1}{2}}}  & \multigate{2}{\Pi} \\
    \lstick{\{I_{i+1}\}}       &  \qw & \ghost{\rotatebox{90}{QCrank }}        & \qw &\ghost{\sum_{\frac{1}{2}}}&\\
    \lstick{\{I_{i-1}\}}       &  \qw & \ghost{\rotatebox{90}{QCrank }}        & \gate{X} & \multigate{1}{\sum_{\frac{1}{2}}}&\ghost{\Pi}  &  \measuretab{Z} &  \rstick{ \{ G^2_{x_i}\}}\\
    \lstick{\{I_{i+1}\}}       &  \qw & \ghost{\rotatebox{90}{QCrank }}        & \qw &\ghost{\sum_{\frac{1}{2}}}\\
}
\caption{\textbf{Quantum circuit computing $G_x^2$.} A serialized image is stored in 4 copies on 4 data qubits of \qcrank.}
\label{fig:gradx2-circ}
\end{figure}
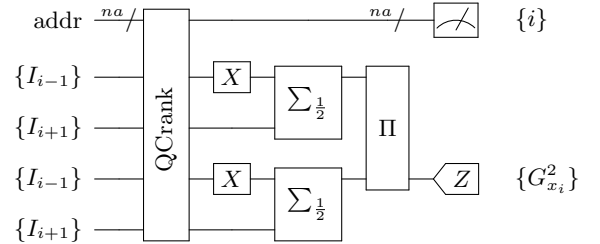

The circuit (\cref{fig:gradx2-circ}) assumes four copies of the input image are serialized into 1D vectors with appropriate index offsets ($i \pm 1$), encoded via \qcrank\ on $4 + n_a$ qubits where $L = 2^{n_a}$ is the total number of pixels. In pre-processing, gray-pixel values are linearly transformed to satisfy $I(i,j) \in [-1, 1]$, and boundary conditions are handled by replicating edge pixels.

The \eh\ protocol implements: (i) negation of the $i-1$ values with X-gates, (ii) the horizontal gradient $G_x$ computed twice in parallel using the weighted sum operator $\Sigma_{1/2}$, and (iii) multiplication of the two gradient copies using the product operator $\Pi$. The expectation value of the output data qubit, measured along with all address qubits, enables reconstruction of the 2D gradient image during post-processing. A fully expanded version of this circuit, showing all gates, is provided in~\cref{fig:circ_sqrGrad}, specifically for short sequences where \( L = 4 \).


\subsubsection{Edge Detection}
\label{sec:circ_edge}

The most complex application implements a complete computer vision algorithm for edge detection by combining 2D gradient computations with thresholding. Edge detection is based on the inequality:
\begin{equation}
G_x^2 + G_y^2 - T > 0,
\label{eq:edge_detect}
\end{equation}
where $T$ is a user-defined threshold.

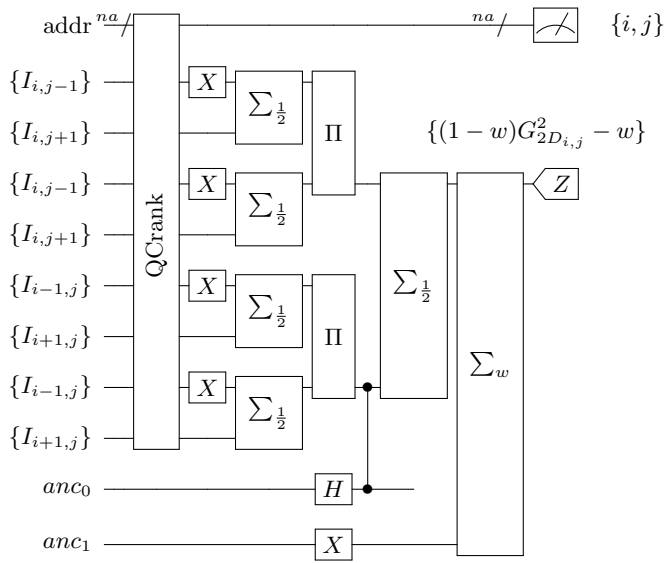
\begin{figure}[t!]
\centering
\hspace{2mm}
\Qcircuit @C=0.4em @R=1em {
    \lstick{\text{addr}}  & ^{na}{/} \qw   &\qw&   \multigate{8}{\rotatebox{90}{QCrank }}  &\qw     &\qw   &\qw   &\qw   &\qw    & ^{na}{/} \qw    & \meter & & \rstick{ \{{i,j}\}}\\
    \lstick{\{I_{i,j-1}\}}      &\qw   &  \qw & \ghost{\rotatebox{90}{QCrank }}        &  \gate{X}  & \multigate{1}{\sum_{\frac{1}{2}}}  & \multigate{2}{\Pi} \\
    \lstick{\{I_{i,j+1}\}}      &\qw   &  \qw & \ghost{\rotatebox{90}{QCrank }}        & \qw &\ghost{\sum_{\frac{1}{2}}}&&&\rstick{ \{ (1-w) G_{2D_{i,j}}^2 - w \} }\\
    \lstick{\{I_{i,j-1}\}}      &\qw   &  \qw & \ghost{\rotatebox{90}{QCrank }}        & \gate{X} & \multigate{1}{\sum_{\frac{1}{2}}}&\ghost{\Pi}  & \qw & \multigate{4}{\sum_{\frac{1}{2}}} & \multigate{7}{\sum_{w}} &  \measuretab{Z}  \\
    \lstick{\{I_{i,j+1}\}}      &\qw   &  \qw & \ghost{\rotatebox{90}{QCrank }}        & \qw &\ghost{\sum_{\frac{1}{2}}} \\
    \lstick{\{I_{i-1,j}\}}      &\qw   &  \qw & \ghost{\rotatebox{90}{QCrank }}        &  \gate{X}  & \multigate{1}{\sum_{\frac{1}{2}}}  & \multigate{2}{\Pi} & &\\
    \lstick{\{I_{i+1,j}\}}     &\qw    &  \qw & \ghost{\rotatebox{90}{QCrank }}        & \qw &\ghost{\sum_{\frac{1}{2}}}\\
    \lstick{\{I_{i-1,j}\}}     &\qw    &  \qw & \ghost{\rotatebox{90}{QCrank }}        & \gate{X} & \multigate{1}{\sum_{\frac{1}{2}}}&\ghost{\Pi}  & \control \qw & \ghost{\sum_{\frac{1}{2}}} \\
    \lstick{\{I_{i+1,j}\}}      &\qw   &  \qw & \ghost{\rotatebox{90}{QCrank }}        & \qw &\ghost{\sum_{\frac{1}{2}}}&&& &\\
    \lstick{anc_0}   &\qw& \qw  & \qw  & \qw  & \qw  &  \gate{H}  & \ctrl{-2} &\qw &  &\\
    \lstick{anc_1}   &\qw& \qw  & \qw   & \qw  & \qw &\gate{X} & \qw  & \qw  &\ghost{\sum_{w}}\\
    &\\
}
\caption{\textbf{Quantum circuit used for computing $ G_x^2 + G_y^2-T>0 $}.  A total of $10+n_a$ qubits are required and $1+n_a$  are measured.}
\label{fig:circ-edge}
\end{figure}

The circuit (\cref{fig:circ-edge}) computes the squared gradients in both directions simultaneously, $G_x^2/4$ and $G_y^2/4$, following the method discussed in the previous section. These are subsequently averaged:
\begin{equation}
G_{2D}^2 = \frac{G_x^2 + G_y^2}{8}.
\label{eq:g2d}
\end{equation}

The final operation is another weighted average of $G_{2D}^2$ and the value $-1$ using weight $w = T/(8-T)$, resulting in an expectation value:
\begin{equation}
\text{EV} = (1-w)G_{2D}^2 - w,
\label{eq:edge_ev}
\end{equation}
where $w$ controls the user defined detection threshold. The condition $\text{EV} > 0$ from~\cref{eq:edge_ev} is equivalent to the edge detection criterion in~\cref{eq:edge_detect}. By post-selecting all addresses measured on the $n_a$ address qubits under the condition $\text{EV} > 0$, we identify pixels locations belonging to edges obeying~\cref{eq:edge_detect}.

The circuit requires $10 + n_a$ qubits with measurements of $1 + n_a$ qubits to determine edge membership, where 8 qubits are needed for extra copies of data and 2 ancilla qubits support the weighted summation computations. The need for the CZ-gate before gradients are averaged is explained in~\cite{balewski2025ehands}.

%
%
%
%
%

\section{Experimental Evaluation}
\label{sec:results}

This section presents experimental validation of the \mo{}  framework described in Section~\ref{sec:framework}. 
We first describe the experimental methodology, including error sources and problem-dependent calibration procedures, then present results for each of the four applications.


\subsection{Experimental Setup and Methodology}
\label{sec:setup}

The mathematical exactness of \qcrank\ and \eh\ operations, when evaluated as abstract unitary transformations of the state vector, is proven in the Supplementary Material of~\cite{balewski2025ehands}.
This work examines the accuracy of selected experiments performed on real quantum hardware (convolution and squared gradient) and, for larger problems, using an ideal shot-based quantum simulator (DFT and edge detection).

We anticipate two primary sources of error that will impact the accuracy of our experiments:  
(i) \textit{shot noise}, which decreases with the number of shots according to $\sigma_{\text{shot}} \sim \nicefrac{1}{\sqrt{N_{\text{shot}}}}$, and  
(ii) \textit{cumulative gate infidelity}, $\sigma_{\text{hw}}$, present only in real hardware experiments, which increases with the total number of CNOT gates required by the problem after transpilation to the native connectivity of the IBM QPU.

These two types of error are uncorrelated and therefore add in quadrature:
\begin{equation}
\sigma_{\text{tot}} = \sqrt{\sigma_{\text{shot}}^2 + \sigma_{\text{hw}}^2} 
= \sqrt{\frac{A}{N_{\text{shot}}} + B},
\label{eq:noiseModel}
\end{equation}
where $A, B$ depend on the complexity of the quantum circuit and the specific problem being solved, and $B$ depends also on the hardware characteristics (gate fidelity, noise levels). For an ideal simulator, $B = 0$; hence, with a sufficiently large number of shots (and unbounded computational resources), the \qcrank\ and \eh\ framework can achieve arbitrary accuracy.

Our evaluation addresses three key research questions: (i) How accurate are the results at fixed shot count compared to classically computed ground truth? (ii) How does accuracy of results change with problem complexity, circuit size, and QPU choice? (iii) What are the practical limitations and error scaling when implementing these algorithms on current NISQ hardware?

\paragraph{Evaluation procedure.}
All experiments follow a standardized procedure to ensure reproducible quantum-classical comparisons. Expected results are computed via classical numerical methods and serve as the ground truth reference. 
Problem-specific circuits are then simulated in Qiskit or transpiled to the QPU and executed there, with shot counts chosen large enough to keep statistical errors small but non-negligible. We designed experiments such that the $B$-term in Eq.~\ref{eq:noiseModel} dominates the error.

For QPU runs, measured expectation values are scaled during post-processing by a single \textit{attenuation} parameter to recover the dynamic range of the ground truth. 
Finally, the algorithm accuracy is quantified via root mean square error (RMSE) between quantum outputs and classical references.

\begin{figure}[h!]
\centering
\includegraphics[width=0.99\columnwidth]{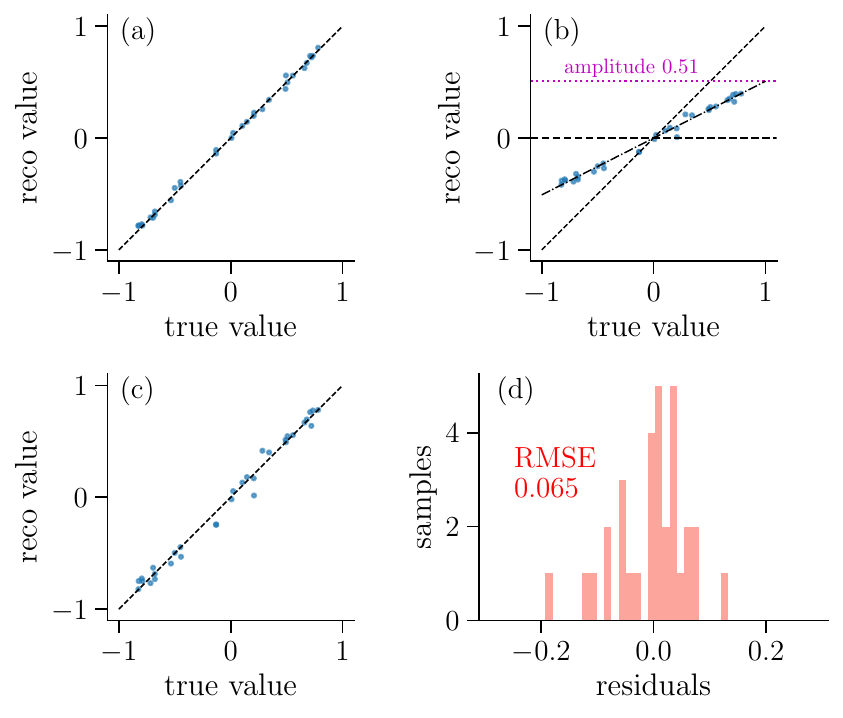}
\caption{\textbf{Post-processing of experimental results for convolution problem}. 
(a) Results from shot-based ideal simulator. Tight correlation along diagonal indicates nearly exactly computed convolution.
(b) Raw IBM Pittsburgh output for the same number of shots has reduced dynamic range by about half. (c) Experimental results rescaled by the factor of 1.96. (d) Histogrammed residuals of data from (c) are used to compute RMSE.  }
\label{fig:2D_RMSE}
\end{figure}

\paragraph{Post-processing calibration.}
Systematic attenuation of the output signal on hardware, primarily due to gate infidelity, is compensated in post-processing by rescaling all measured expectation values for a given problem by a single factor.  
The procedure, illustrated in \cref{fig:2D_RMSE} for the convolution experiment, derives this attenuation factor from the correlation between raw quantum measurements (\cref{fig:2D_RMSE}b) and ground truth (\cref{fig:2D_RMSE}a).  
In this example, multiplying the raw results by $1.96$ yields corrected experimental values (\cref{fig:2D_RMSE}c), whose residuals are then used for RMSE computation.

This single-parameter adjustment mirrors standard QPU calibration practices. With 30–60 measured output values per experiment, removing one degree of freedom minimally impacts data correlation, allowing us to assert that results remain essentially unbiased.  
Without this step, experimental outputs would track the ground truth but with reduced amplitude, making similarity harder to interpret.

\paragraph{Experimental configurations.}
Table~\ref{tab:qcrank_size} summarizes our four experiments, ranging from a simple convolution of 2 real-valued lists of lengths 32 (7 qubits, 164 CNOT gates, 32,000 shots) to a demanding gradient-based edge detection computation on a gray-valued image of 25,000 pixels (24 circuits, each with 20 qubits, over 8,000 CNOT gates, and 30 million shots).
Shot counts were chosen so that statistical errors were much smaller than the signal amplitude yet remained visible in plots.

Hardware runs were executed on IBM quantum processors (Pittsburgh, Kingston, Aachen) with Pauli twirling enabled for error mitigation.  
For the squared gradient and edge detection tasks, a single monolithic circuit was infeasible—either exceeding the practical CNOT count limits for the hardware or requiring qubit and shot counts impractical for simulation.  
For those cases, the input image was tiled, processed via multiple smaller circuits, and the results stitched together during post-processing, as indicated in Table~\ref{tab:qcrank_size}.

\begin{table}[h!]
\vspace{6pt}
\begin{center}
\caption{Configurations of experiments}
\begin{tabular}{c|c|c|c|c}
\textbf{problem:} & \textbf{Conv 1D} & \textbf{DFT} & $\mathbf{G_x^2}$ & \textbf{Edge det} \\
\hline
input type &\multicolumn{2}{c|}{real-value lists} &\multicolumn{2}{c}{gray image}\\
input size & 2 $\times$ 32 & 11 $\times$ 512 & 32 $\times$ 32 & 192 $\times$ 128 \\
num circ & 1 & 3 \ddag  & 64 & 24\\
address qubits & 5 & 9 & 4 & 10\\
total qubits & 7 & 20 & 8 & 20\\
CZ gates & 33 & 5642 & 69 & 8207\\
transp. CZ's & 164 & 5642 & 252 & 8207\\
backend & Pittsburgh \dag & ideal & Aachen \dag & ideal\\
IBM scaling & 1.96 & - & 2.9 & -\\
shots/circ & 32k & 1M & 100k & 30M\\
circuit & \cref{fig:circ-conv} & \cref{fig:circ-DFT} & \cref{fig:gradx2-circ} & \cref{fig:circ-edge}\\
results & \cref{fig:1dConv} & \cref{fig:gravWave-meas} & \cref{fig:gradx2-meas} & \cref{fig:bacteria-edge}\\
\hline
\end{tabular}\\
\label{tab:qcrank_size}
\dag) IBM quantum hardware with enabled Pauli Twirling\\
\ddag) 1 circuit yielded DFT results for 5 selected frequencies
\end{center}
\end{table}


\subsection{Convolution Results}
\label{sec:results_conv}

\begin{figure}[htb]
\centering
\includegraphics[width=0.95\columnwidth]{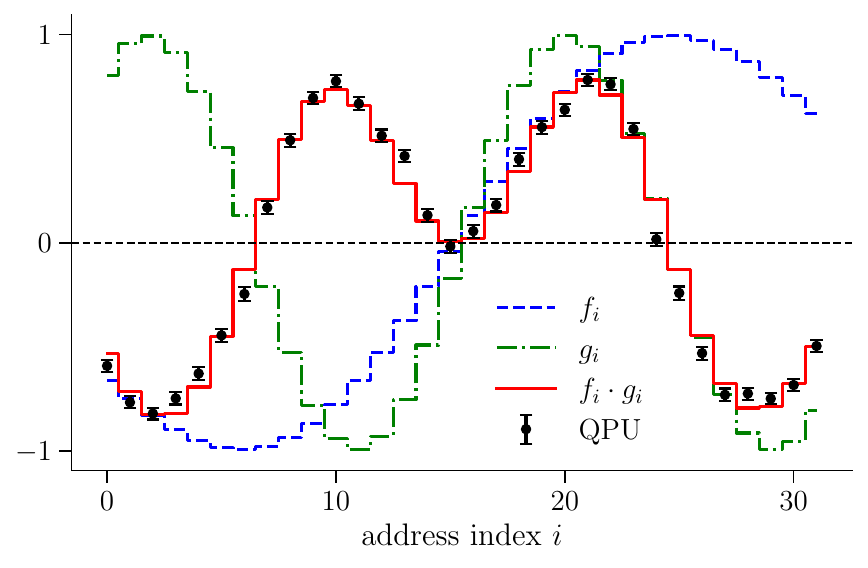}
\vspace{-12pt}
\caption{\textbf{Computation of convolution using IBM Pittsburgh QPU}. Two input functions are shown as dashed lines, ground truth output is solid red, and measurements are shown as points with statistical errors.}
\vspace{-12pt}
\label{fig:1dConv}
\end{figure}

\Cref{fig:1dConv} presents results from IBM Pittsburgh for the convolution of two arbitrarily chosen functions $f_i, g_i$, each discretized into 32 bins.
The experiment used the circuit shown in \cref{fig:circ-conv} with 7 qubits and 32,000 shots, achieving an RMSE of 0.065 after post-processing calibration.

To study the resilience of this method to the noise inherent in NISQ devices, we repeated  experiments several times with different lengths of input sequences on two types of IBM QPUs, as shown in Table~\ref{tab:hw_std}. 
The number of shots was scaled with the sequence length to ensure that the statistical error of the measurements remained constant and negligible. 
A clear degradation in precision is observed with an increasing number of required two-qubit gates.
The performance of the two investigated QPUs was very similar.

We conclude it is feasible to reliably compute convolution on IBM NISQ QPUs
(available to us in Spring of 2025) as long as the transpiled circuit requires below 200 CNOT gates.

\begin{table}[h!]
\vspace{6pt}
\begin{center}  
\caption{Standard deviation of convolution results on IBM quantum hardware.}
\begin{tabular}{c|c|c|c|c}
\textbf{seq len} & \textbf{shots} & \textbf{2q gates \dag} & \textbf{Kingston} & \textbf{Pittsburgh} \\
\hline
16  & 16,000 & 69& 0.043 & 0.047 \\
32  & 32,000 & 164& 0.087 & 0.065 \\
64 & 64,000  & 342& 0.140 & 0.126 \\
\hline
\end{tabular}\\
\label{tab:hw_std}
\dag) two-qubit gate counts after transpilation
\end{center}
\end{table}


\subsection{DFT Results}
\label{sec:results_DFT}

This experiment demonstrates how the integral of convolution of two functions can be computed on a QPU by simply not measuring address qubits. To better reflect real-world use cases, we apply the DFT calculation to a synthetic gravitational wave signal.

\begin{figure}[ht!]
\centering
\includegraphics[width=0.99\columnwidth]{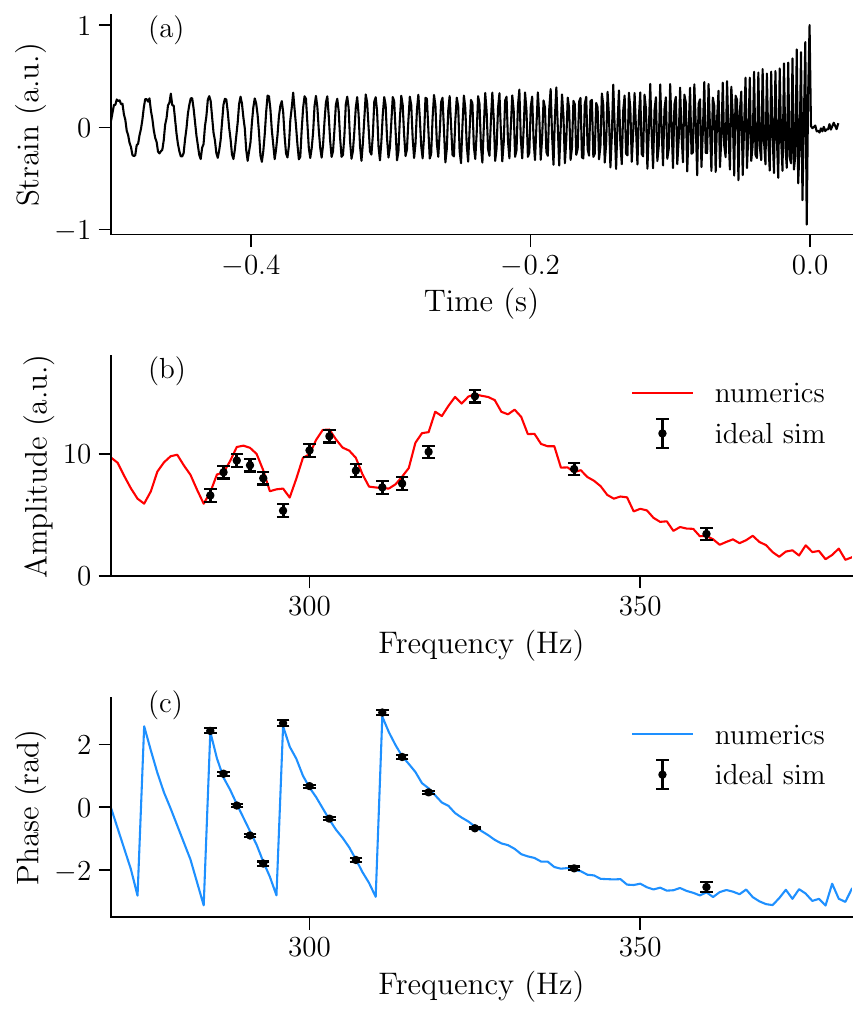}
\caption{\textbf{Fourier transform of gravitational wave strain for 15 selected frequencies}. (a) Normalized input being a sequence of 512 real values. (b) Amplitude and (c) phase of Fourier transform of the input. The results from ideal simulator are shown as points with statistical errors.}
\label{fig:gravWave-meas}
\end{figure}

\paragraph{Synthetic gravitational-wave chirp signal.}
As a test case, we simulate the dimensionless gravitational-wave \textit{strain} $h(t)$ generated during the final inspiral phase of a compact binary coalescence~\cite{Isoyama2021}. 
The strain amplitude evolves with time $t$ approximately as:
\begin{equation}
h(t) \sim f(t)^{2/3} \cos(2\pi f(t)t + \phi(t)),
\end{equation}
where $f(t)$ is the instantaneous gravitational-wave frequency, and $\phi(t)$ is the evolving phase.
We discretize $h(t)$ into $N = 512$ time steps, as shown in \cref{fig:gravWave-meas}(a). The ground truth for the corresponding  amplitudes and phases as functions of probe frequencies are shown in \cref{fig:gravWave-meas}(b) and (c) as solid lines.

The objective of the DFT experiment is to choose $k=5$ probe frequencies $\omega_k$ and construct a single 20-qubit quantum circuit (\cref{fig:circ-DFT}). \qcrank\ encodes the strain $h_i$ time series on the lowest qubit and the 5 pairs of modulation lists $\{\cos(\omega_k t_i), \sin(\omega_k t_i)\}$ on 10 qubits. The top 9 qubits provide the common indexing for all 11 input lists. The result of 10 multiplication operators is measured, yielding the 5 pairs of $I(\omega_k), Q(\omega_k)$ values. By not measuring the address qubits, we let the QPU compute the average \textit{in-situ}. The conversion from $I, Q$ to amplitude and phase is done in classical post-processing.

We repeated the quantum circuit simulation 3 times, for a total of 15 choices of probe frequency. 
The results, shown as points with errors in \cref{fig:gravWave-meas}(b) and (c), agree very well with the classical ground truth. 
The deviations are purely due to the shot noise and can be made arbitrary smaller if larger  number of shots is used, following ~\cref{eq:noiseModel}.
Both amplitude and phase measurements demonstrate high fidelity across the selected frequency range, validating the quantum algorithm's correctness.
Since the circuit requires about 6000 CNOT gates before transpilation, it was clear from IBM's published CNOT gate fidelities that execution on IBM QPUs was not practical.

\subsection{Squared Gradient Results}
\label{sec:results_grad}
The objective is to compute the squared directional gradient for a gray-pixel image using circuit shown in Fig.~\ref{fig:gradx2-circ}.
For the $32 \times 32$ pixel test image ($L = 1024$ pixels), hardware constraints necessitated dividing the image into 64 strips of size $1 \times 16$ pixels each. Each strip required $n_a = \lceil \log_2(16) \rceil = 4$ address qubits, resulting in circuits  with 8 total qubits and 252 CNOT gates after transpilation for the heavy-hex topology of IBM QPUs.

\begin{figure}[h!]
\centering
\includegraphics[width=0.9\columnwidth]{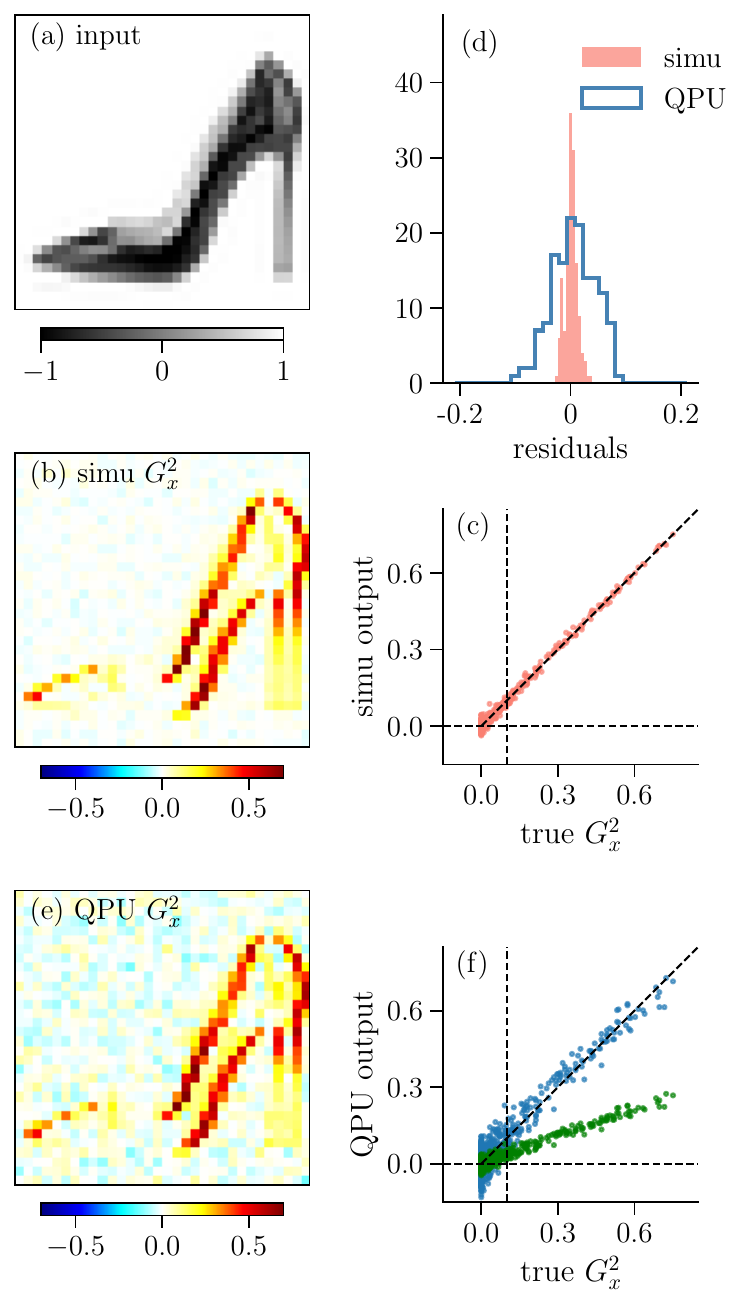}
\vspace{-12pt}
\caption{\textbf{Squared X-Gradient for an image.} 
(a) Input image with 1024 gray-scale pixels. (b) Ideal simulation output and (c) its correlation  with ground truth $G_x^2$. (d) Histogram of residuals when true $G_x^2 > 0.1$. (e,f) Analogous to (b,c): $G_x^2$ results from IBM Aachen. Green and blue data on (f) show correlation before and after scaling by the factor of 2.9.}
\label{fig:gradx2-meas}
\end{figure}

The 64 circuits were executed on IBM Aachen using 100,000 shots each. \Cref{fig:gradx2-meas} demonstrates strong visual correspondence between the quantum results (panel e) and classical gradient computation (panel b). This experiment confirms image processing capabilities on NISQ hardware, highlighting the viability of quantum approaches for computer vision tasks. The implementation generates coherent gradient maps that reveal key structural features of the input image. 

Although the calibration factor (2.9) exceeds that of simpler circuits, this predictable and correctable scaling behavior indicates consistent performance. The subdivision approach effectively manages larger images, illustrating how algorithmic strategies can enhance the capabilities of NISQ devices. The agreement observed for significant gradient values (\(G_x^2 > 0.1\)) affirms that quantum circuits can reliably detect important image features, essential for edge detection and related applications. These findings lay the groundwork for more advanced quantum image processing algorithms as hardware capabilities improve.

\subsection{Edge Detection Results}
\label{sec:results_edge}
\begin{figure}[ht!]
\includegraphics[width=.9\linewidth]{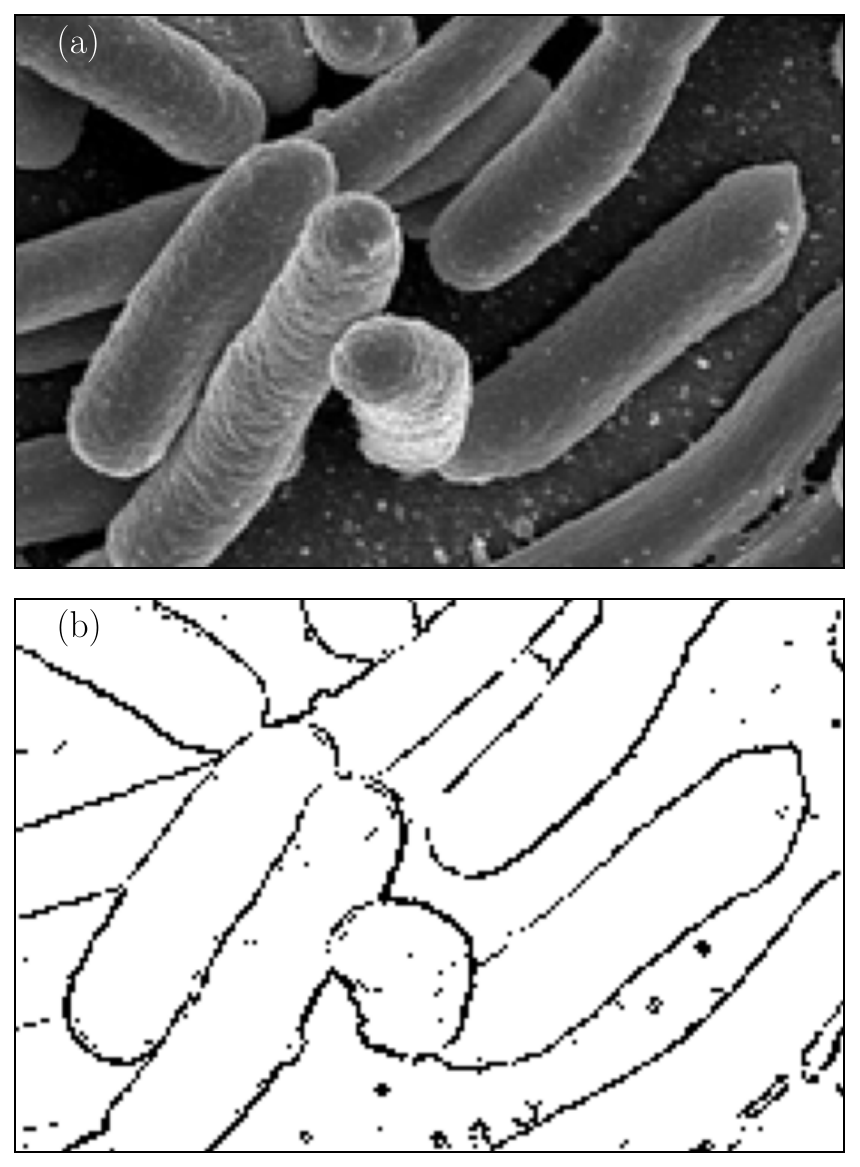}
\caption{\textbf{Edge detection with quantum circuit}. (a) Input gray-scale image, (b) simulated output with binary edges selected.}
\label{fig:bacteria-edge}
\end{figure}

Tagging pixels on the edges of photographed objects based on the large magnitude of the two-dimensional gradient is possible with the circuit shown in Fig.~\ref{fig:circ-edge}. We demonstrate this by processing an image of bacteria, as shown in \cref{fig:bacteria-edge}a. The algorithm correctly identifies boundaries and internal structural features while suppressing noise, resulting in a clean binary edge map shown in panel b.

For practical reasons, the $192 \times 128$ pixel image was subdivided into 24 tiles with 1024 pixels, processed individually and stitched in post-processing. 
Each tile required $n_a = 10$ address qubits. The circuit consists of a total of 20 qubits, has over 8000 CNOT gates, and 30 million shots were needed for the statistical precision demonstrated in the output image.

This experiment demonstrates the feasibility of implementing complete computer vision algorithms using the \qcrank+\eh\ framework. 
The ability to perform parallel gradient computation and thresholding operations in a single quantum circuit represents an important step toward practical quantum image processing systems.

\section{Conclusion and Future Work}
\label{sec:conclusion}

This work presents \mo, the \qcrank+\eh\ framework for quantum data processing and demonstrates its practical implementation on NISQ devices. 
By leveraging the shared EVEN encoding between \qcrank\ data representation and \eh\ polynomial transformations, we construct shallow, non-variational circuits that implement convolution, discrete-time Fourier transform, squared gradient computation, and edge detection.

Our systematic progression from 7-qubit hardware implementations to 20-qubit simulations demonstrates both current NISQ capabilities and near-term scalability. 
Hardware experiments on IBM quantum processors achieved meaningful accuracy for circuits with up to approximately 200 transpiled CNOT gates, while larger computations required tiling strategies to manage accuracy degradation from cumulative gate infidelity.
The linear scaling properties of \eh\ polynomial circuits and the end-to-end pipeline—from classical input encoding through quantum transformation to interpretable outputs—address a critical gap between theoretical quantum algorithms and practical implementations.

Current limitations include substantial shot requirements (32K–30M) and accuracy degradation for high gate counts or high-degree polynomials.
While our implementations do not yet achieve quantum advantage over classical methods, they establish feasibility benchmarks and provide a foundation for future development.
Beyond research contributions, this work offers educational insights into the practical challenges of NISQ algorithm development.

Future research directions include scaling studies as hardware improves, extensions to machine learning applications, and hybrid classical-quantum approaches. 
The polynomial computation foundation enables numerous extensions including nonlinear filtering and advanced computer vision operations.

While long-term practical utility is expected to arise from fault-tolerant quantum computers, current NISQ devices play an important role as stepping stones—enabling early exploration of quantum data processing algorithms and deeper understanding of the requirements for achieving quantum advantage in data-intensive applications.
\section{Acknowledgments}

This research was supported by the U.S. Department of Energy (DOE) under Contract No.~DE-AC02-05CH11231, through the Office of Science, Office of Advanced Scientific Computing Research (ASCR) Exploratory Research for Extreme-Scale Science and Accelerated Research in Quantum Computing (ARQC).
This research used resources of two DOE user facilities: the National Energy Research Scientific Computing Center (NERSC) located at Lawrence Berkeley National Laboratory, operated under Contract No.~DE-AC02-05CH11231, and the Oak Ridge Leadership Computing Facility, operated under Contract No.~DE-AC05-00OR22725.

\bibliography{0_main.bbl}

\end{document}